\documentclass[12pt,pra,aps,amssymb,amsmath,tightenlines]{revtex4}

\bibstyle{unsrt}
\begin{document}
\tighten
\draft


\title{Combinatorics and field theory}

\author{Carl~M.~Bender$^{1}$,
Dorje~C.~Brody$^{2}$, and Bernhard~K.~Meister$^{3}$}

\affiliation{${}^{1}$Department of Physics, Washington University,
St. Louis MO 63130, USA \\ ${}^{2}$Blackett Laboratory, Imperial
College, London SW7 2BZ, UK \\ ${}^{3}$Department of Physics, Renmin
University of China, Beijing 100872, China}

\date{\today}
\input{psfig.sty}

\begin{abstract}
For any given sequence of integers there exists a quantum field
theory whose Feynman rules produce that sequence. An example is
illustrated for the Stirling numbers. The method employed here
offers a new direction in combinatorics and graph theory.
\end{abstract}

\maketitle

\section{Introduction}
\label{s1}

In quantum field theory graphs are used to represent the terms in
a diagrammatic perturbation expansion, whereby one associates with
each graph a numerical amplitude. It is known \cite{1} that there
is an elegant formal construction for the set of all graphs in
terms of the exponential of a derivative operator. For example,
\begin{eqnarray}
Z(\epsilon,g)=\exp({1\over2!}\epsilon d^{2}/dx^{2})\exp({1\over4!}
gx^{4})|_{x=0} \nonumber
\end{eqnarray}
is the generating function for the set of all 4-vertex vacuum
diagrams, connected and disconnected, in which the line amplitude is
$\epsilon$ and the vertex amplitude is $g$. Here, for simplicity, we
consider field theories in zero-dimensional space so that Feynman
integrals become trivial and are merely the product of the line
amplitudes.

In analogy with statistical mechanics we refer to the sum of all
vacuum diagrams as the {\it partition function}. The power series
expansion of a generic partition function contains both connected
and disconnected graphs. If we are interested only in connected
graphs, then we consider instead the {\it free energy} $F=-\ln Z$;
the coefficients of the power series expansion of $F$ represent the
sum of the symmetry numbers of just the connected graphs \cite{2},
where the symmetry number of a graph is defined as the reciprocal
of the number of ways in which the graph can be turned into itself
by permuting the lines or vertices. These are the basic rules for
graphs in field theories. Using these rules, we would like to find
field theories whose diagrammatic expansions correspond to graphs
that are meaningful in combinatorics. A simple but intriguing
example is the field theory of partitions.

\section{Field theory of partitions}

The partition of an integer $n$ is the set of all distinct ways
to represent $n$ as a sum of positive integers smaller than or
equal to $n$. The number of elements in the partition of $n$ is
designated $P_{n}$. Defining $P_{0}=1$, the first few $P_{n}$
are $1,~1,~2,~3,~5,~7,~11,\cdots$. In the case of partitioning
of labelled objects, the number of partitions is given by the Bell
numbers $\{ B_{n}\}=1,~1,~2,~5,~15,~52,~203,\cdots$. The labelled
partitions can also be grouped into classes characterised by the
Stirling numbers $S(n,k)$, which count the number of partitions
of $n$ labelled objects into $k$ groups \cite{3}. Specifically, we
have $S(1,1)=1$, $S(2,1)=S(2,2)=1$, $S(3,1)=1$, $S(3,2)=3$,
$S(3,3)=1$, and so on. Clearly, if we sum $S(n,k)$ over $k$, we
recover the Bell numbers: $B_{n}=\sum_{k}S(n,k)$.

There is an elementary field theory whose Feynman rules produce
precisely the labelled partitions into groups \cite{4}. This is
given by the potential energy $V(x)=g(e^{x}-1)$ and the
{\it  line insertion operator} $D=\epsilon d/dx$ for a
grouping variable $g$ and a line amplitude $\epsilon$:
\begin{eqnarray}
Z(\epsilon,g)\ =\ \exp\left(\epsilon\frac{d}{dx}\right)
\exp\left[g(e^{x}-1)\right]\Biggm|_{x=0} \ =\
\sum_{n=0}^{\infty}\frac{\epsilon^{n}}{n!}
\left( \sum_{k=1}^{n}S(n,k)g^{k}\right)\ . \nonumber
\end{eqnarray}
If we set $g=1$, then this reduces to the field theory of labelled
partitions (the Bell numbers). Note that the line insertion operator
in the above example is given by a first order derivative operator
$d/dx$. Graphically, such propagators correspond to lines having
only one end. This is a strange kind of line; ordinarily, a line has
two ends. However, in the field theory, we can have generalised
lines having multiple ends.

\section{Generalised partitions}

Consider a general field theory that includes all $n$-point vertices
as well as all generalised lines having $m$ ends. This leads to the
expression
\begin{eqnarray}
Z(L,V)\ =\
\exp\left(\epsilon\sum_{m=1}^{\infty}
{L_m\over m!}{d^m\over dx^m}\right)
\exp\left(g\sum_{n=1}^{\infty}{V_n\over n!}x^n\right)
\Biggm|_{x=0}\ , \nonumber
\end{eqnarray}
which represents the set of all vacuum diagrams, connected and
disconnected, constructed from $n$-point vertices whose amplitudes
are $V_n$ and generalised lines having $m$ legs whose amplitudes are
$L_m$. If we expand this expression as a formal series in powers of
$\epsilon$, then the coefficient of $\epsilon^{n}$ is the sum of the
symmetry numbers of all graphs having $n$ lines, and if we expand
this expression as a series in powers of $g$, then the coefficient
of $g^n$ is the sum of the symmetry numbers of all graphs having $n$
vertices. We note that in general these formal power series are
divergent because the number of graphs grows like a factorial.

In particular, if we set $g=1$ and $L_{m}=V_{n}=1$, then we
obtain the following beautiful result on generalised partitions:
\begin{eqnarray}
Z(\epsilon)\ =\ \exp\left( e^{\epsilon\frac{d}{dx}}-1\right)
\exp\left( e^{x}-1\right) \Biggm|_{x=0} \ =\
\sum_{n=0}^{\infty} \frac{\epsilon^{n}}{n!}B_{n}^{2}\ .
\nonumber
\end{eqnarray}
The graphs contributing to this expression, up to order
$\epsilon^{4}$, are shown in Fig. 1 below.

\begin{figure*}[th]
   \psfig{file=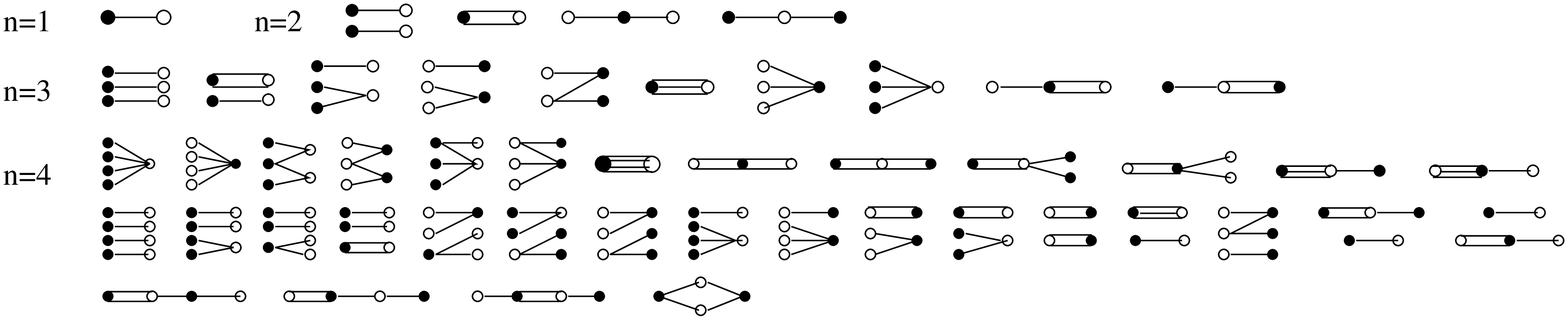,width=15cm,angle=0}
\vskip .25cm
\caption{Graphs in a theory whose Feynman rules allow for
$n$-point vertices ($n=1,~2,~3,\cdots$) and $m$-legged lines
($m=1,~2,~3,\cdots$). If the vertex amplitudes are all unity
and the $m$-legged line amplitude is $\epsilon^{m}$, then the
generating function $Z(\epsilon)$ for the graphs has a Taylor
expansion for which the coefficient of $\epsilon^{n}$ is
$B_{n}^{2}/n!$. Thus, the number of labelled graphs
of order $n$ is the square of the $n$th Bell number.}
\end{figure*}

It is interesting that the graphs of generalised lines in the above
figure have formal resemblance to the graphs in twistor diagrams.
These generalised lines are, however, quite natural in the context
of quantum field theory. For example, the strong-coupling expansion
of the Lagrangian for quantum chromodynamics are known to involve
these multilegged propagators. Note that the number of these graphs
form a sequence $1,~1,~4,~10,~33,\cdots$. We do not know how to
generate this sequence, however, if we label the vertices in Fig. 1,
then the number of graphs for generalised labelled partitions are
given by the square of the Bell numbers.

\section{Topology numbers}

We have illustrated how the Feynman rules in quantum field theory
can naturally be associated with ideas in combinatorics. The
structures we introduced above are, however, quite primitive in
the sense that we worked only in zero-dimensional space, and we have
not introduced Fermion lines, which give rise to directed graphs.
Despite such simplicity in the underlying field theory, the
corresponding graphical expansions are already quite intricate. This
suggests that, by introduction of additional structures (or perhaps
even without such extensions) we might be able to obtain partition
functions that would generate unknown, crucially important integer
sequences such as topology numbers or partially ordered sets (posets).

Although we do not know how this might be achieved, in the following
we would like to sketch the line of thinking involved in such
problems. Here, let us consider the labelled topologies. The $n$-th
topology number can be represented by unlabelled transitive graphs
of $n$ nodes. By transitivity, we mean if $\alpha$ and $\beta$ are
related and $\beta$ and $\gamma$ are related, then $\alpha$ and
$\gamma$ also have to be related. Some of the graphs are shown in
Fig. 2.

\begin{figure*}[th]
   \psfig{file=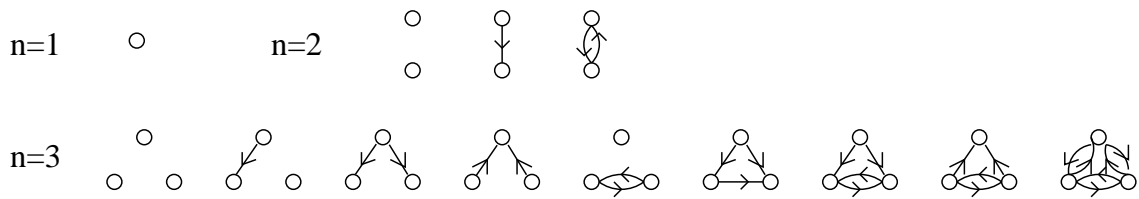,width=15cm,angle=0}
\vskip .25cm
\caption{Connected and disconnected graphs representing topology
numbers: 1, 1, 3, 9, 33, 139, 718, 4535, $\cdots$. Only eight
terms of the sequence are known.}
\end{figure*}

Since disconnected graphs can be obtained by exponentiating the
connected ones in the sense noted above, let us consider only
the connected ones (free energy). The labelled connected
topologies are then given by the sum of the symmetry numbers of
the connected transitive graphs. Because we are interested in
the symmetry numbers, all the transformations of the graph that
preserve the symmetry numbers are allowed. Thus, we may simplify
the above graphs in the following manner. The first step is to
replace the double lines in Fig. 2 by single `Bosonic' lines. The
connected graphs obtained after this replacement are shown in Fig. 3.

\begin{figure*}[th]
   \psfig{file=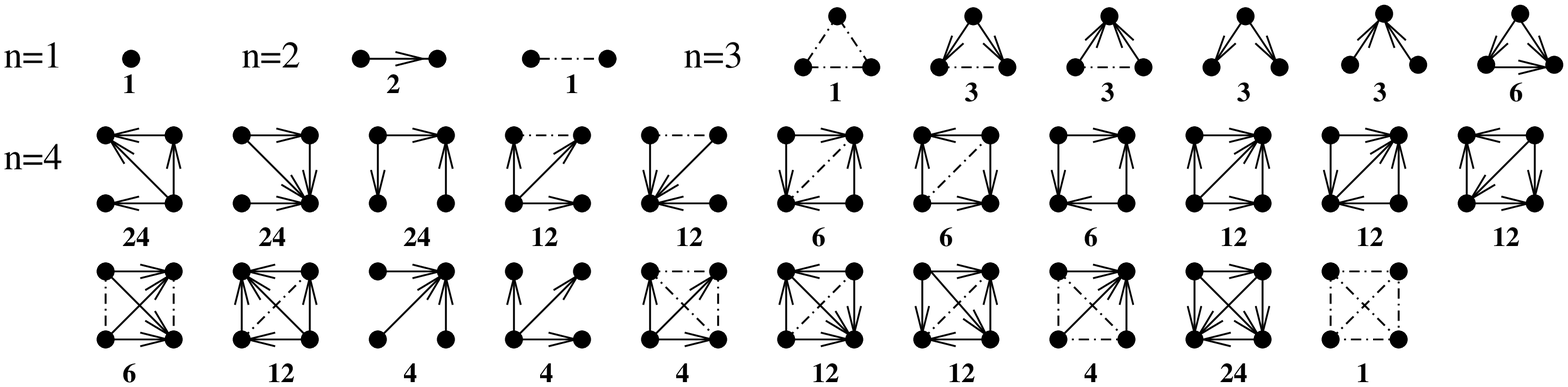,width=15cm,angle=0}
\vskip .25cm
\caption{Graphs representing connected topologies, after replacing
the double arrows in Fig. 2 by single Bosonic lines. The numbers
are the associated symmetry numbers times $n!$.
}
\end{figure*}

The second step is to `squash' all the Bosonic lines, namely, any
graph having $n$ vertices joined by Bosonic lines can be squashed
into a single $n$-vertex. The symmetry number is preserved by
associating the factor of $1/n!$ to each $n$-vertex. The final step
is then to lift the transitivity in the sense that, if $\alpha$
is related to $\beta$ and $\beta$ is related to $\gamma$, then the
redundant relation joining $\alpha$ and $\gamma$ by a `Fermion'
line (arrow) can be removed. It can easily be seen that this
simplification does not alter the symmetry numbers. After these
transformations, the simplified graphs appear to be identical to
the connected posets. An example for $n=4$ in Fig. 3 is shown in
Fig. 4.

\begin{figure*}[th]
   \psfig{file=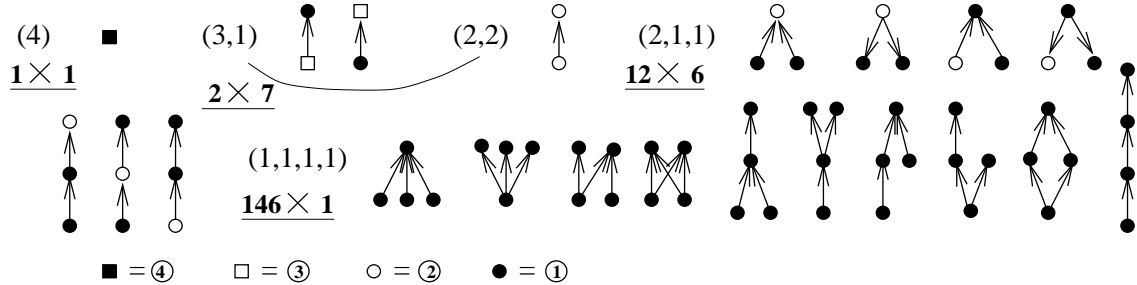,width=15cm,angle=0}
\vskip .25cm
\caption{The 21 graphs for $n=4$ in Fig. 3 reduce to these graphs,
which are just the connected posets. In particular, the squashing
of Bosonic lines gives five partitions of 4. The corresponding
Stirling numbers are 1, 7, 6, 1 respectively for one, two, three,
and four groups to partition 4. The number of connected labelled
posets, on the other hand, are given by 1, 2, 12, 146, and so on.
Therefore, we deduce that the number of connected labelled
topologies for $n=4$ is given by $1\times1 + 2\times7 + 12\times6
+ 146\times1 = 233$, without counting the symmetry numbers
indicated in Fig. 3.
}
\end{figure*}

The number of graphs corresponding to Fig. 4, if we label the
vertices, can be determined by counting the associated symmetry
numbers. However, we can deduce this number without such a
procedure, because the graphs in Fig. 4 are in fact precisely the
connected posets, and the Stirling numbers tells us how many ways
we can partition the number in the group, as stated in the figure
caption. Therefore, if we let
$\{ d_{n}\}=1,~2,~12,~146,~3060,~101642,\cdots$ denote connected
labelled posets and $\{ t_{n}\}=1,~3,~19,~233,~4851,~158175,\cdots$
denote the connected labelled topologies, we deduce the following
formula:
\begin{eqnarray}
t_{n}\ =\ \sum_{k=1}^{n} S(n,k) d_{k}\ .
\nonumber
\end{eqnarray}
By exponentiating the foregoing arguments in the sense noted
earlier, we can show that the same identity via Stirling numbers
also holds for disconnected graphs, a formula known in combinatorics
\cite{3}.

In the above discussion on topology numbers our line of thinking in
simplifying the problem is motivated by field theoretic ideas. The
challenging problem is to find a field theory whose Feynman rules
produce graphs corresponding to those in Fig. 4. This may be
achieved by introducing Fermions, Wick ordering, and so on. We
note that only the first 14 terms in the sequences $\{ d_{n}\}$,
$\{ t_{n}\}$ are known.


\begin{enumerate}

\bibitem{1} C.M.~Bender and W.E.~Caswell, {\em J.~Math.~Phys.}
\textbf{19}, 2579 (1978).

\bibitem{2} C.M.~Bender, F.~Cooper, G.S.~Guralnik, D.H.~Sharp,
R.~Roskies, and M.L.~Silverstein, {\em Phys.~Rev.~D} \textbf{20},
1374 (1979).

\bibitem{3} L.~Comtet, {\it Advanced Combinatorics} (Reidel,
Dordrecht, 1974).

\bibitem{4} C.M.~Bender, D.C.~Brody, and B.K.~Meister,
{\em J. Math. Phys}, \textbf{40}, 3239 (1999).

\end{enumerate}
\end{document}